\documentclass[conference,letter,10pt]{IEEEtran}

\usepackage[utf8]{inputenc} 
\usepackage[T1]{fontenc}
\usepackage{url}
\usepackage{ifthen}
\usepackage[cmex10]{amsmath} 

\usepackage{refcheck}
\usepackage{graphicx}
\usepackage{amsmath}
\usepackage{amssymb}
\usepackage{graphics}
\usepackage{latexsym}
\usepackage[dvips]{epsfig}
\usepackage[nospace]{cite}
\usepackage{subfigure}
\usepackage{epstopdf}
\usepackage{mathdots}
\usepackage{fullpage}

\newtheorem{Lemma}{\bf Lemma}
\newtheorem{Corollary}{\bf Corollary}
\newtheorem{Proposition}[Lemma]{\bf Proposition}
\newtheorem{Theorem}{\bf Theorem}

\newtheorem{Remark}{Remark}

\def\E{{\rm \mathbf  E}}

\def\X{{\mathcal{X}}}
\def\Y{{\mathcal{Y}}}
\def\Z{{\mathcal{Z}}}

\newcommand{\ml}[2]{\mathcal{L}\left(#1  \!\!  \to  \!\!   #2\right)}

\allowdisplaybreaks

\begin{document}

\sloppy

\title{Strengthened Information-theoretic Bounds on the Generalization Error}

 \author{
   \IEEEauthorblockN{Ibrahim Issa, Amedeo Roberto Esposito, Michael Gastpar}
  \IEEEauthorblockA{EPFL\\
    \{ibrahim.issa,amedeo.esposito,michael.gastpar\}@epfl.ch,} 
 }

\pagenumbering{gobble} 

 \maketitle

\begin{abstract} 
The following problem is considered: given a joint distribution $P_{XY}$ and an event $E$, bound $P_{XY}(E)$ in terms of $P_XP_Y(E)$ (where $P_XP_Y$ is the product of the marginals of $P_{XY}$) and a measure of dependence of $X$ and $Y$. Such bounds have direct applications in the analysis of the generalization error of learning algorithms, where $E$ represents a large error event and the measure of dependence controls the degree of overfitting. Herein, bounds are demonstrated using several information-theoretic metrics, in particular:  mutual information, lautum information, maximal leakage, and $J_\infty$. The mutual information bound can outperform comparable bounds in the literature by an arbitrarily large factor.
\end{abstract}

\section{Introduction}

One of the main challenges in designing learning algorithms is guaranteeing that they generalize well~\cite{False,FalsePositive,PreventFalse,
LearnabilityStability}. The analysis is made especially hard by the fact that, in order to handle large data sets, learning algorithms are typically adaptive. 
A recent line of work initiated by Dwork \emph{et al.}~\cite{StatisticalValidity,Holdout,algoStability} shows that differentially private algorithms provide generalization guarantees. More recently, Russo and Zou~\cite{RussoZou}, and Xu and Raginsky~\cite{RaginskyGeneralization}, provided an information-theoretic framework for this problem, and showed that the mutual information (between the input and output of the learning algorithm) can be used to bound the generalization error, under a certain assumption. Jiao \emph{et al.}~\cite{jiao2017dependence} and Issa and Gastpar~\cite{ComputableBounds}  relaxed this assumption and provided new bounds using new information-theoretic measures. 

The aforementioned papers focus mainly on the expected generalization error. In this paper, we study instead the \emph{probability} of an undesirable event (e.g., large generalization error in the learning setting). In particular, given an event $E$ and a joint distribution $P_{XY}$, we bound $P_{XY}(E)$ in terms of $P_XP_Y(E)$ (where $P_XP_Y$ is the product of the marginals of $P_{XY}$) and a measure of dependence between $X$ and $Y$. 

Bassily \emph{et al.}~\cite{littleinfo} and Feldman and Steinke~\cite{calibratingvariance} provide a bound of this form, where the measure of dependence is mutual information, $I(X;Y)$. We present a new bound in terms of mutual information, which can outperform theirs by an arbitrarily large factor. Moreover, we prove a new bound using lautum information (a measure introduced by Palomar and Verd{\'u}~\cite{lautum}). We demonstrate two further bounds using maximal leakage~\cite{CompSecQuantLeakage,MyMaximalLeakage} and $J_\infty(X;Y)$ (which was recently introduced by Issa and Gastpar~\cite{ComputableBounds}). One advantage of the latter two bounds is that they can be computed using only a partial description of the joint $P_{XY}$, hence they are more amenable to analysis.

\section{Main Results}

Let $P_{XY}$ be a joint probability distribution on alphabets $\X \times \Y$, and let $E \subseteq \X \times \Y$ be some (``undesirable'') event. We want to bound $P_{XY}(E)$ in terms of $P_{X}P_Y (E)$ (where $P_X P_Y$ is the product of the marginals induced by the joint $P_{XY}$) and a measure of dependence between $X$ and $Y$.

\subsection{KL Divergence/Mutual Information Bounds}

Consider the following intermediate problem: let $P$ and $Q$ be two probability distributions on an alphabet $\Z$, and let $E \subseteq \Z$ be some event. We will bound $P(E)$ in terms of $Q(E)$ and $D(P||Q)$. Then by replacing $P$ by $P_{XY}$ and $Q$ by $P_X P_Y$, we get a bound for our desired setup in terms of the mutual information $I(X;Y)=D(P_{XY} || P_X P_Y)$.

\begin{Proposition} \label{lemma:boundsimple}
Given $q \in (0,1)$, define $f_q: [q,1] \rightarrow \mathbb{R}_+$ as $f_q(p)=D(p||q)$. Then, $f_q(p)$ is a strictly increasing function of $p$. Given any event $E$ and pair of distributions $P$ and $Q$ with $D(P||Q) \leq \log \frac{1}{Q(E)}$, 
\begin{align} \label{eq:boundsimple}
P(E) \leq f^{-1}_{Q(E)} \left(D(P||Q)\right).
\end{align}
In particular, given an event $E \subseteq \X \times \Y$ and a joint distribution $P_{XY}$ satisfying $I(X;Y) \leq \log \frac{1}{P_X P_Y (E)}$, \begin{align} \label{eq:boundsimpleMI}
P_{XY}(E) \leq f^{-1}_{P_XP_Y(E)} \left(I(X;Y)\right).
\end{align}
\end{Proposition}

\begin{IEEEproof}
Note that $ \frac{df_q(p)}{dp} = \log \left( \frac{p}{q} \frac{1-q}{1-p} \right) >0$ for $p>q$, hence $f_q(p)$ is strictly increasing. 
Moreover, the range of $f_q(p)$ is $[0,\log(1/q)]$, so~\eqref{eq:boundsimple} is well defined.

If $P(E) \leq Q(E)$, then~\eqref{eq:boundsimple} holds trivially since $f_{Q(E)}^{-1} (D(P||Q)) \geq Q(E)$ by the definition of $f$. Otherwise, if $P(E) > Q(E)$, then $f_{Q(E)} \left(P(E) \right) = D \left( P(E) || Q(E) \right) \leq D(P||Q)$, where the second inequality follows from the data processing inequality. Since $f_q$ is strictly increasing, then so is $f_q^{-1}$. Hence $P(E) \leq f^{-1}_{Q(E)} \left(D(P||Q)\right)$.
\end{IEEEproof} 

The bound above is tight in the following sense. Let $g: [0,1] \times \mathbb{R}_+  \rightarrow [0,1]$ be such that, given any alphabet $\Z$ and event $E \subseteq \Z$, and any two distributions $P$ and $Q$ on $\Z$, $P(E) \leq g \left( Q(E), D(P||Q) \right)$. Then $ g \left( Q(E), D(P||Q) \right) \geq f^{-1}_{Q(E)} \left(D(P||Q)\right)$ if $D(P||Q) \leq \log \frac{1}{Q(E)}$. This is true since given any tuple $(\Z,P,Q,E)$ such that $D(P||Q) \leq \log \frac{1}{Q(E)}$, there exists $(\Z',P',Q',E')$ such that $Q(E')=Q(E)$, $D(P||Q) = D(P' || Q')$, and~\eqref{eq:boundsimple} holds with equality. In particular, choose $\Z'=\{0,1\}$, $E' = \{1\}$,  $Q' \sim \mathrm{Ber}(Q(E))$,
and $P' \sim \mathrm{Ber} \left( f_{Q(E)}^{-1} \left(D (P||Q) \right) \right)$. 

However, there is no closed form for the bound in~\eqref{eq:boundsimple}. The following corollary provides an upper bound in closed form:

\begin{Corollary} \label{corr:bdexp} Given $q \in (0,1/2]$, define $g_q(y) := \log^2(2)+\left(\log(1-q)+y\right)\left(-\log(q)-y\right)$ and 
$ \hat{f}_q: [0, -\log(q)) \rightarrow \mathbb{R}_+$ as follows:   
\begin{align*} \label{eq:upperboundfunc} 
& \hat{f}_q(y)  =  \\
& \frac{2\log^2(2)+ \! \left(\log(1 \! -q)+y\right)\log\frac{(1-q)}{q}+ \! (\log 4) \sqrt{g_q(y)} }{\log^2\left((1-q)/q \right)+\log^2(2)}.
\end{align*}
Then, $\hat{f}_q(y)$ is concave and non-decreasing in $y$. Moreover, given any event $E$ and pair of distributions $P$ and $Q$ with $D(P||Q) \leq \log \frac{1}{Q(E)}$, 
\begin{align}
P(E) \leq \hat{f}_{Q(E)} \left(D(P||Q) \right).
\end{align}
In particular, given an event $E \subseteq \X \times \Y$ and a joint $P_{XY}$ satisfying $I(X;Y) \leq \log \frac{1}{P_X P_Y (E)}$, \begin{align} \label{eq:corrMI}
P_{XY}(E) \leq \hat{f}_{P_XP_Y(E)} \left(I(X;Y)\right).
\end{align}
\end{Corollary}

\begin{figure*}[t]
\hrulefill
\begin{align}
& \lim_{q \rightarrow 0} \lim_{D(P||Q) \rightarrow 0}
\frac{\tilde{f}_q(D(P||Q))}{\hat{f}_q(D(P||Q))} \notag  \\
& \stackrel{(a)}=
\lim_{q \rightarrow 0} \frac{ \left(2\log^2(2)+q_1q_2 +(2\log2)\sqrt{\log^2(2)-q_2\log(q)} \right) \log(1/q) }{ \left(q_1^2+\log^2(2) \right) \log (2) }  \notag \\
& = \lim_{ q \rightarrow 0} 
\frac{ \left( 2\log^2(2)+\log^2(1-q)- \log(q)\log(1-q)+(2\log2) \sqrt{\log^2(2)-\log(1-q)\log(q)} \right)\log(1/q)  }{ \left(\log^2(1-q) + \log^2(q)-2\log(q)\log(1-q)+ \log^2(2) \right) \log(2) }  \notag \\
& \stackrel{(b)} = \lim_{q \rightarrow 0} \frac{4\log^2(2)\log(1/q)}{\left(\log^2(q)+\log^2(2)\right)\log(2)}  \notag \\
& = 0,
\end{align}
where in (a) $q_1 = \log \frac{1-q}{q}$ and $q_2 = \log(1-q)$, and (b) follows from the fact that $\lim_{q \rightarrow 0} \log(q) \log(1-q) = 0$.

\hrulefill
\end{figure*}

\begin{IEEEproof}
Since $g_q(y)$ is concave in $y$ and the square root is concave and non-decreasing, $\sqrt{g_q(y)}$ is concave in $y$; hence $\hat{f}_q(y)$ is concave in $y$. To show that it is non-decreasing, consider the derivative (ignoring the positive denominator):
\begin{align*}
\frac{d \hat{f}_q(y)}{d y} = \log \frac{1-q}{q} + \log(4) \frac{-2y -\log(q(1-q))}{2 \sqrt{g_q(y)}}.
\end{align*} 
For $y \in [0, -\frac{1}{2} \log(q(1-q))]$, both terms are non-negative (the first is non-negative since $q \leq 1/2$). For $y \in [-\frac{1}{2} \log(q(1-q)), -\log(q)]$, the numerator of the second term is negative and decreasing, and the denominator is positive and decreasing. Hence, it achieves its minimum for $y = -\log(q)$. Since the minimum $  \frac{d \hat{f}_q(y)}{d y} \bigg|_{y = -\log(q)} = 0$, we get that $ \frac{d \hat{f}_q(y)}{d y} \geq 0$ for $y \in [0,-\log(q)]$.

Now, let $p:=P(E)$ and $q:=Q(E)$. Then we can rewrite the inequality $D(p||q) \leq D(P||Q)$ as 
\begin{align} \label{eq:expand}
\! -\log(1-q)+ \! p \log \left(\frac{1-q}{q} \right) -h(p) \leq D(P||Q),
\end{align} where $h(.)$ is the binary entropy function (in nats).
Upper-bounding $h(p) \leq (\log 4) \sqrt{p(1-p)}$ , we get
\begin{align*}
 - & \log(1 \! -q) \!+ \! p \log \!\frac{1 \!- \!q}{q} \! - \!(\log 4)\! \sqrt{p(1\!-\!p)}  \leq D(P||Q).
\end{align*}
For ease of notation, let $y:=D(P||Q)$ and $\tilde{g}(p)$ be the left-hand side. Then, 
\begin{align}
\frac{d \tilde{g}}{dp} = \log \left( \frac{1-q}{q} \right) - (\log 4) \frac{1-2p}{\sqrt{p(1-p)}}.
\end{align}
Hence, there exists $p_0$ such that $\tilde{g}$ is decreasing on $[0,p_0]$ and increasing on $[p_0,1]$. Therefore, $\tilde{g}(p)= y$ admits at most two solutions, say $p_1 < p_2$, and $\tilde{g}(p) \leq y \Rightarrow p \leq p_2$.  It remains to solve
\begin{align}
\! p \log \frac{1-q}{q} - \! \log \left(1-q \right) - \! (\log 4)\sqrt{p(1-p)} = y.
\end{align} Let $q_1 = \log \frac{1-q}{q}$, and $q_2 = \log(1-q)$.
 We get 
\begin{align}
 (pq_1 - q_2-y)^2 & = p(1-p) \log^2(4), \iff \notag \\
p^2 \left(q_1^2+\log^2(4) \right) & -2p\left(2\log^2(2)+q_1(q_2+y) \right) \notag \\
& +(q_2+y)^2  = 0.  \label{eq:quadratic}
\end{align}
The discriminant of~\eqref{eq:quadratic} is given by \begin{align*}
& \frac{\Delta}{4} = \! \left(2\log^2(2) \! +q_1(q_2+y) \! \right)^2 \!\! - \! (q_1^2+\! \log^2(4))(q_2 \! +y)^2 \\
&  = (q_2 \! +y)(4 q_1 \! \log^2(2)-(\log^2(4))(q_2 \! +y)) \! + \! 4 \log^4(2) \\
&  = \left(4\log^2(2)\right)\left(\log^2(2)+(q_2+y)(q_1-q_2-y)\right)  \geq 0,
\end{align*} 
where the inequality follows from the fact that $q_1-q_2-y = -\log(q)-y \geq 0$. Hence, the larger root of~\eqref{eq:quadratic} is given by $\hat{f}_q(p)$, as desired.
\end{IEEEproof}

\subsubsection{Comparison with existing bound}

It has been shown~\cite[Lemma 3.11]{calibratingvariance}\cite[Lemma 9]{littleinfo} that 
\begin{align} \label{eq:boundlit}
P(E) \leq \frac{D(P||Q)+\log(2)}{\log\left(1/Q(E)\right)}.
\end{align}
The bound in Corollary~\ref{corr:bdexp} can be arbitrarily smaller than~\eqref{eq:boundlit}. That is, let $\tilde{f}_{Q(E)}(D(P||Q))$  be the left-hand side of~\eqref{eq:boundlit} and consider the calculation shown at the top of the next page.

Moreover, one can derive a family of bounds in the form of~\eqref{eq:boundlit} using the Donsker-Varadhan characterization of KL. In particular,
\begin{align} \label{eq:KLvarchar}
\! D(P||Q) = \!\!\!\!\! \sup_{ \substack{ f: \Z \rightarrow \mathbb{R},  \E_Q[e^f] < +\infty}} \!\!\!\!\!\! \left\lbrace \E_P[f] - \log \E_Q[e^f] \right\rbrace. \!\!
\end{align}
Now, let $f = \beta \mathbb{I} \{ z \in E \}$ for some $\beta >0$, where $\mathbb{I}\{ \}$ is the indicator function. After rearranging terms, we get
\begin{align} \label{eq:DVKL}
P(E) \leq \frac{D(P||Q) + \log \left(1+(e^\beta-1)Q(E) \right)}{\beta}.
\end{align}
Choosing $\beta = \log(1/Q(E))$, we slightly improve~\eqref{eq:boundlit} by replacing $\log(2)$ with $\log(2-Q(E))$. 
In fact, we can solve the infimum over $\beta >0$ of the right-hand side of~\eqref{eq:DVKL}. In particular, by~\cite[Lemma 2.4]{ConcentrationInequalities}, the infimum is given by $\ell^{\star -1}(D(P||Q))$, where $\ell^\star$ is the convex conjugate of\footnote{Lemma 2.4 of~\cite{ConcentrationInequalities} assumes $\ell''(0)=0$, but the proof goes as is for $\ell''(0) \geq 0$, which is the case here.} $\ell(\beta) = \log(1+(e^\beta-1)Q(E))$, and $\ell^{\star -1}(y) = \inf \{t: \ell^\star(t) > y\}$. It turns out that $\ell^\star: \mathbb{R}_+ \rightarrow \mathbb{R}_+$ is given by
\begin{align} \label{eq:ellstar}
\ell^\star(t) = \begin{cases}
0, & 0 \leq t < Q(E), \\
D(t||Q(E)), & Q(E) \leq t \leq 1, \\
+\infty, & t>1.
\end{cases}
\end{align}
Now, $P(E) \leq \inf\{t: \ell^{\star}(t) > D(P||Q)\}$. Hence, for $D(P||Q) = 0$, $P(E) \leq \inf (Q(E),+\infty) = Q(E)$. By noting that $\ell^\star(1)=\log(1/Q(E))$, we get for any $D(P||Q) > \log(1/Q(E))$, $P(E) \leq \inf(1,+\infty) = 1$. Finally, for $D(P||Q) \in (0, \log(1/Q(E))]$, we get $P(E) \leq \{t \in [Q(E),1]: D(t||Q(E)) > D(P||Q) \}$, which is equal to $t^\star \in [Q(E),1]$ satisfying $D(t^\star ||Q(E)) = D(P||Q)$.
That is, the bound derived from~\eqref{eq:ellstar} exactly recovers Proposition~\ref{lemma:boundsimple}. 

Furthermore, we could compare with the ``mutual information bound'' of Russo and Zou~\cite{RussoZou}, and Xu and Raginsky~\cite{RaginskyGeneralization}. In particular, by considering $f = \beta \left( \mathbb{I} \{ z \in E \}-Q(E) \right)$ for $\beta \in \mathbb{R}$ in~\eqref{eq:KLvarchar}, we get
\begin{align*}
D(P||Q) & \geq \!  \beta (P(E) \!\! -Q(E)) \! - \log \! \E_Q \!\! \left[e^{\left( \mathbb{I} \{ Z \in E \}-Q(E) \right)} \right] \\
& \geq \beta (P(E)-Q(E)) - \beta^2/8,
\end{align*} 
where the second inequality follows from the fact that $(\mathrm{Ber}(q)-q)$ is $\frac{1}{4}$-subgaussian (which is true for any random variable whose support has length 1).  Since the above inequality holds for any $\beta \in \mathbb{R}$, we get
\begin{align} \label{eq:Raginskybound}
P(E) \leq Q(E) + \sqrt{\frac{D(P||Q)}{2}}.
\end{align}
Evidently, Corollary 1 can outperform~\eqref{eq:Raginskybound} since (for finite $D(P||Q)$) $\lim_{Q(E) \rightarrow 0} \hat{f}_{Q(E)} \left( D(P||Q) \right) = 0$, whereas the right-hand side of~\eqref{eq:Raginskybound} goes to $\sqrt{D(P||Q)/2}$. In Figure 1, we plot the 3 bounds (equations \eqref{eq:expand},~\eqref{eq:boundlit}, and~\eqref{eq:Raginskybound}) for a given range of interest: small $Q(E)$, and relatively small $D(P||Q)$, e.g., proportional to $-\log(1-Q(E))$. 
\begin{figure}[htp]
\centering
\subfigure[$D(P||Q)=-2\log(1-Q(E))$]{\includegraphics[scale=0.36]{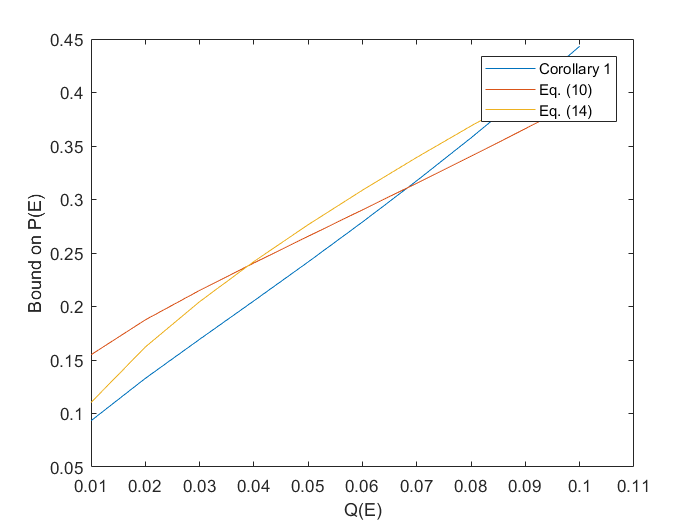}}
\subfigure[$D(P||Q)=-3\log(1-Q(E))$]{\includegraphics[scale=0.36]{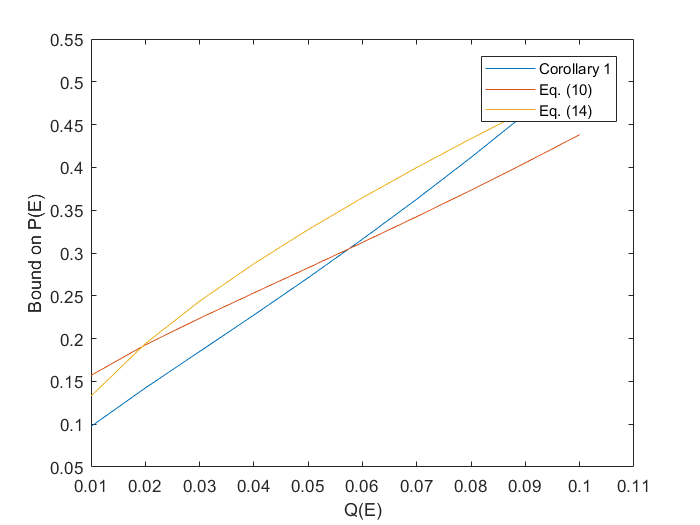}}
\subfigure[$D(P||Q)=-4\log(1-Q(E))$]{\includegraphics[scale=0.36]{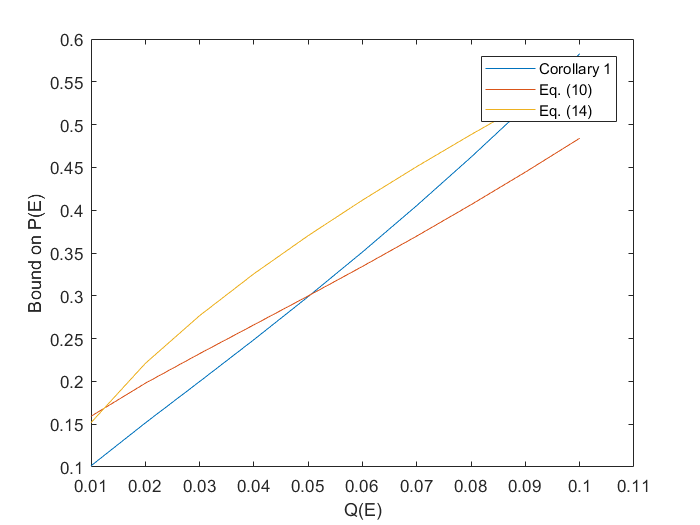}}
\caption{Comparison of bounds}
\end{figure}
\begin{Remark}
Given the form of the 3 bounds, one might expect that~\eqref{eq:Raginskybound} outperforms the other two for large values of $D(P||Q)$. This is in fact not true because the range of interest for the right-hand sides is restricted to $[0,1]$. For instance, for small $Q(E)$ and $D(P||Q) = -\log(Q(E))/2$, the bound in~\eqref{eq:Raginskybound} is trivial ($>1$), and the other two bounds are strictly less than 1.
\end{Remark}

\subsubsection{Bound using $D(Q||P)$/Lautum Information}

By considering the data processing inequality $D(q||p) \leq D(Q||P)$, we can bound $p$ in terms of $q$ and $D(Q||P)$. 
\begin{Theorem} \label{prop:DQP}
Given any event $E$ and a pair of distributions $P$ and $Q$, if $P(E) \leq 1/2$, then
\begin{align*}
P(E) \leq 1-e^{-h \left(Q(E) \right) - D(Q||P)}.
\end{align*}
In particular, given an event $E \subseteq \X \times \Y$ and a joint distribution $P_{XY}$ with $P_{XY}(E) \leq 1/2$, \begin{align} \label{eq:lautum}
P_{XY}(E) \leq 1-e^{-h \left(P_XP_Y(E) \right) - L(X;Y)},
\end{align}
where $L(X;Y):=D(P_XP_Y||P_{XY})$ is \emph{the lautum information}~\cite{lautum}.
\end{Theorem}
\begin{IEEEproof} Set $p=P(E)$ and $q=Q(E)$.
As in~\eqref{eq:expand}, we can rewrite $D(q||p) \leq D(Q||P)$ as 
\begin{align} \label{eq:expand2}
q \log \left(\frac{1-p}{p} \right)-\log(1-p) -h(q) \leq D(Q||P).
\end{align}
Since $p \leq 1/2$ (by assumption), we can drop the first term of the left-hand side. Rearranging the inequality then yields Theorem~\ref{prop:DQP}. 
\end{IEEEproof}

Moreover, we can derive a family of bounds similar to~\eqref{eq:DVKL} by considering the Donsker-Varadhan representation of  $D(Q||P)$:
\begin{align} \label{eq:KLvarchar2}
\!\! D(Q||P) = \!\!\!\!\! \sup_{ \substack{ f: \Z \rightarrow \mathbb{R},  \E_P[e^f] < +\infty}} \!\!\!\!\!\! \left\lbrace \E_Q[f] - \log \E_P[e^f] \right\rbrace. \!\!
\end{align}
Now, let $f = -\beta \mathbb{I} \{ z \in E \}$ for some $\beta >0$. Then after rearranging terms, we get for any $\beta >0$,
\begin{align} \label{eq:DVKL2}
P(E) \leq \frac{1-e^{-D(Q||P)-\beta Q(E)}}{1-e^{-\beta}}.
\end{align}

\subsection{Maximal Leakage Bound}
The bounds presented so far in~\eqref{eq:corrMI} and~\eqref{eq:lautum} do not take into account the specific relation of $P_{XY}$ and $P_X P_Y$ as a joint distribution and its marginal. Indeed, they are applications of a more general bound that can be applied to an arbitrary pair of distributions (Corollary~\ref{corr:bdexp} and Theorem~\ref{prop:DQP}). The following bound does not fall under this category, i.e., it only applies to pairs of distributions forming a joint and marginal.

\begin{Theorem} \label{prop:MaxL}
Given $\alpha \in [0,1]$, finite alphabets $\X$ and $\Y$, a joint distribution $P_{XY}$ and an event $E \subseteq \X \times \Y$ such that for all $y \in \Y$, $P_X(E_y) \leq \alpha$ where $E_y := \{ x: (x,y) \in E \}$, then 
\begin{align}
P_{XY}(E) \leq \alpha \exp \left\lbrace \ml{X}{Y} \right\rbrace,
\end{align}
where $\ml{X}{Y} = \log \sum_{y \in \Y} \max_{\substack{x : \\ P_X(x) >0}} P_{Y|X} (y|x)$ is the maximal leakage.
\end{Theorem}  
\begin{Remark}
The bound holds more generally but we restrict our attention to finite alphabets to make the presentation of the proof simple. 
\end{Remark}

\indent {\emph{Proof:}}
Fix $y \in \Y$ satisfying $P_Y(y)>0$, and consider the pair of distributions $P_{X|Y=y}$ and $P_X$:
\begin{align*}
D_\infty (P_{X|Y=y} || P_X) & = \sup_{A \subseteq \X} \frac{P_{X|Y=y}(A)}{P_X(A)} \\
& = \max_{x: P_{X|Y}(x|y) > 0} \frac{P_{X|Y}(x|y)}{P_X(x)}.  
\end{align*} where the equalities follow from~\cite[Theorem 6]{RenyiKLDiv}.
Hence,
\begin{align*}
P_{X|Y=y}(E_y) & \leq \alpha \max_{x: P_{X|Y}(x|y) > 0} \frac{P_{X|Y}(x|y)}{P_X(x)} \\
& = \alpha  \max_{x: P_{X|Y}(x|y) > 0} \frac{P_{Y|X}(y|x)}{P_Y(y)}.
\end{align*}
Now,
\begin{align*}
P_{XY}(E) & = \E_Y \left[ P_{X|Y=y}(E_y) \right]\\
& \leq \alpha \sum_{y: P_Y(y) > 0} \max_{x: P_{X|Y}(x|y) > 0} P_{Y|X}(y|x) \\
& \stackrel{(a)} = \alpha \sum_{y: P_Y(y) > 0} \max_{x: P_X(x)>0} P_{Y|X}(y|x) \\
& = \alpha \sum_{y \in \Y} \max_{x: P_X(x)>0} P_{Y|X}(y|x)
\end{align*}
where (a) follows from the following (readily verifiable) facts:
\begin{align*}
P_Y(y) > 0 & \text{ and } P_{X|Y}(x|y) > 0 \Rightarrow  P_X(x) > 0, \\
P_Y(y) > 0 & \text{ and } P_{X|Y}(x|y) = 0 \Rightarrow P_{Y|X}(y|x) = 0. \quad \hspace{1mm} \hfill \blacksquare
\end{align*}

One advantage of the bound of Theorem~\ref{prop:MaxL} is that it depends on a partial description of $P_{Y|X}$ only. Hence, it is simpler to analyze than the mutual information bounds. Moreover, for fixed $P_X$, the bound is convex in $P_{Y|X}$. In the next subsection, we present a bound with similar properties.

\subsection{$J_\infty$-Bound}
\begin{Theorem} \label{prop:Jinf}
Given $\alpha \in [0,1/2]$, finite alphabets $\X$ and $\Y$, a joint distribution $P_{XY}$ and an event $E \subseteq \X \times \Y$ such that for all $y \in \Y$, $P_X(E_y) \leq \alpha$ where $E_y := \{ x: (x,y) \in E \}$, then 
\begin{align}
 P_{XY}(E)  \leq \alpha \left( 2(1-\alpha) J_\infty(X;Y) +1 \right),
\end{align}
where $J_\infty(X;Y) = \frac{1}{2}  \sum_{y \in \Y} \left( \max_x P_{Y|X}(y|x) - \min_x P_{Y|X}(y|x) \right) $~\cite{ComputableBounds}.
\end{Theorem}  
\begin{IEEEproof}
The theorem follows from Theorem 1 and Corollary 1 of~\cite{ComputableBounds}. In particular, following the same proof steps as in~\cite{ComputableBounds}, one can show that for any function\footnote{In~\cite{ComputableBounds}, the authors consider $X=(X_1,\cdots,\X_n)$, $\Y=\{1,2,\cdots,n\}$, and $f(X,Y)=X_Y$. Nevertheless, the proof of~\eqref{eq:genf} remains the same.} $f: \X \times \Y \rightarrow \mathbb{R}$, \begin{align} \label{eq:genf}
& \big| \E_{P_{XY}} [f(X,Y)] - \E_{P_X P_Y} [f(X,Y)] \big| \leq \\
& \left( \max_y \E_{P_X} [|f(X,y)-\mu_y|] \right) J_\infty(X;Y), \notag
\end{align}
where $\mu_y := \E_{P_X} [f(X,y)]$. Now, set $f(x,y) = \mathbb{I} \{ (x,y) \in E \}$. Then, $\E_{P_{XY}} [f(X,Y)] = P_{XY} (E)$, $\E_{P_X P_Y} [f(X,Y)] = P_XP_Y(E) \leq \alpha$, and $\E_{P_X} [f(X,y)] = P_X(E_y)$. Moreover,
\begin{align*}
\E_{P_X}[ | f(X,y) - P_X(E_y)|] & = 2 P_X(E_y) \left( 1-P_X(E_y) \right) \\
&  \leq \alpha(1-\alpha), 
\end{align*} 
where the last inequality follows from the assumption that $P_X(E_y) \leq \alpha \leq \frac{1}{2}$. Then, it follows from~\eqref{eq:genf} that
\begin{align}
P_{XY}(E)- P_X P_Y(E) \leq 2 \alpha (1-\alpha) J_\infty(X;Y).
\end{align}
The theorem follows by noting that $P_X P_Y(E) \leq \alpha$.
\end{IEEEproof}

\bibliographystyle{IEEEtran}
\bibliography{IEEEabrv,database}

\end{document}